\documentstyle[12pt,aasms4]{article}
\newcommand{\etal}{et~al.}

\begin{document}
\onecolumn

\title{
Classifications of the Host Galaxies of Supernovae 
}

\author{Sidney van den Bergh}
\affil{Dominion Astrophysical Observatory, Herzberg Institute of
Astrophysics, National Research Council, 5071 West Saanich
Road, Victoria, British Columbia, V9E 2E7, Canada 
(email: sidney.vandenbergh@nrc.ca)}

\centerline{and}

\author{Weidong Li and Alexei V. Filippenko}
\affil{
Department of Astronomy, 601 Campbell Hall, University of California,
Berkeley, CA 94720-3411 (email: wli@astro.berkeley.edu, alex@astro.berkeley.edu)}

\newpage

\begin{abstract}
 
   Classifications on the DDO system are given for the host galaxies of 177
supernovae (SNe) that have been discovered since 1997 during the course of the
Lick Observatory Supernova Search with the Katzman Automatic Imaging Telescope.
Whereas SNe~Ia occur in all galaxy types, it is found, at a high level of
statistical confidence, that SNe~Ib, Ic, and II are strongly concentrated in
late-type galaxies. However, attention is drawn to a possible exception
provided by SN 2001I. This SN~IIn occurred in the E2 galaxy UGC 2836, which was
not expected to harbor a massive young supernova progenitor.

\end{abstract}


\section{Introduction}

   Supernova statistics require a large data base for which the selection
effects are well understood. Such a sample is provided by the 177 supernovae
(SNe) discovered since 1997 by the 75-cm Katzman Automatic Imaging Telescope
(KAIT) during the course of the Lick Observatory Supernova Search (LOSS;
Treffers et al. 1997; Li et al. 2000; Filippenko et al. 2001).  The data base
examined in the present investigation extends from SN 1997bs (1997 Apr. 15) to
2002as (2002 Jan. 30). Both radial velocities and modern spectroscopic
classifications are available for almost all of these objects.  A complete
listing of the Lick Observatory data on all of these discoveries is given at
http://astron.berkeley.edu/$\sim$bait/kait.html.  

A drawback of this data set, however, is that the images of host (parent)
galaxies of SNe obtained with KAIT have a rather coarse scale, short
exposures (25--30 s), and were taken in typical seeing of $\sim 2''$. These data
were therefore supplemented by the images of the same galaxies obtained with
the Palomar 1.2-m Schmidt telescope on 103aO emulsion during the first Palomar
Observatory Sky Survey (POSS I), and by POSS II (Reid et al. 1991) images in
the red on IIIaF emulsion and in the blue-green on IIIaJ emulsion.  (At
the time of writing IIIaJ digitized data were not yet publicly available for
all galaxies in the Survey.) Inspection of these images made it possible to
classify each of the host galaxies of a supernova on the DDO system (van den
Bergh 1960a,b,c).  The somewhat underexposed KAIT images are particularly
well adapted to the examination of the high surface-brightness central regions
of some galaxies. On the other hand they are not very suitable for the
classification of galaxies of below-average surface brightness.  Such objects
can, however, be classified very well on the longer-exposure images obtained in
the POSS I and POSS II. 

Due to the effects of seeing and image resolution the nuclear bulges of some
galaxies appear larger in POSS-I images than they do in the blue-green POSS-II
images. Another factor that has to be taken into account when making
classifications is that bulge size is overestimated in the red POSS-II images,
compared to its size in blue images. This is so because most bulges are
intrinsically red.  Furthermore, spiral structure (which is blue) is suppressed
in red images.  As a result there is a systematic tendency to classify galaxies
too early in red images. Finally, the high signal-to-noise ratio in digitized
IIIaJ images of the POSS II allows one to see spiral structure much more clearly
in the POSS II than in the POSS I.  These images are therefore particularly
useful for the luminosity classification of spiral galaxies. At the time of
writing digitized IIIaJ images were only available for 76\% of the galaxies
contained in the present survey. 

In assigning DDO classification types to the host galaxies of SNe the
information available from each of the four (three) sources of images was taken
into account. The classifications so obtained are collected in Table 1.  In
this table uncertain values are followed by a colon. A ``t" in some
classifications indicates evidence for tidal interactions. All the redshifts
and SN classifications in this table were taken from the KAIT web site.

\section{Comparisons with Other Classifications}
   
   Many of the present classifications were made on small images.  It is
therefore of interest to compare these classifications with those that Sandage
\& Tammann (1981; hereafter S\&T) have published for 27 of the present galaxies. Their
classifications were mostly based on inspection of plates obtained with large
reflectors. A caveat is that it is usually not possible to distinguish between
types E and S0 on the images of the host galaxies of SNe that were
employed in the present supernova survey. Another problem is that it is
sometimes difficult to distinguish distant cD galaxies from distant objects of
type Sab. In such cases the presence, or absence, of a swarm of nearby
early-type galaxies can be used to distinguish between probable cD and Sab
galaxies.

A total of 24 objects contained in both the Shapley-Ames Catalog and in the
present data set on host galaxies of SNe could be placed on the sequence
E-Sa-Sb-Sc ($x$ = 0, 1, 2, 3). The other three objects are NGC 788, NGC 6240, and
NGC 3432, which are classified in the present study as E/Sa, Merger, and S
(respectively), while they were S0/Sa, Sa, and Sc (respectively) in the S\&T
classifications (in other words, not all of these classifications fall on the
sequence E-Sa-Sb-Sc).  For the 24 objects that do fall on this sequence from
both studies, we found $<\Delta x> = -0.04 \pm 0.12$ in the sense $x$(DDO) $-$
$x$(S\&T).  This shows that there is no significant systematic difference between
the present Hubble-type assignments and those by S\&T. The standard deviation
of $x$ was found to be 0.58 Hubble classes, so that the standard error for each
the individual DDO and S\&T classifications is probably $\sim$0.4 Hubble
classes.

For 13 of the host galaxies of SNe in the present sample luminosity
classifications were available on both the DDO system and S\&T systems. The
mean difference between these two data sets (in the sense DDO $-$ S\&T) was
0.11 $\pm$ 0.20 luminosity classes, i.e., there is no statistically significant
difference between these two sets of luminosity classifications. The standard
deviation of the differences between individual luminosity classifications was
0.72 luminosity classes. This suggests that the accuracy of each set of
luminosity classifications was probably $\sim$0.5 luminosity classes. It is
concluded that the present set of galaxy classifications is on a system that
is statistically indistinguishable from those employed by van den Bergh
(1960a,b,c) and by S\&T.

\section{Discussion}

   Table 2 gives a compilation of the frequency with which SNe of
different types occur in galaxies with various Hubble classifications. It is of
interest to note that Li et al.  (2001) discussed the selection process of the
LOSS galaxy sample, and showed that it is representative of the general
population of galaxies. The data in this table indicate the following:

\begin{enumerate}

\item{SNe~Ia occur in galaxies of all Hubble types. This result is in agreement
with previous work (e.g., van den Bergh \& Tammann 1991).}

\item{The frequency distributions of SNe~Ia and SNe Ia-pec, as a function of
Hubble type, do not differ at a respectable level of statistical significance.}

\item{SNe~Ib and SNe~Ic (collectively referred to as ``SNe~Ibc'') occur
mostly in late-type (Sbc-Sc) spirals, whereas SNe of type Ia are mainly
observed in early-type (E-Sb) galaxies. A Kolmogorov-Smirnov test shows that
there is only a 0.6\% probability that the SNe~Ia and SNe~Ibc in the present
sample were drawn from the same frequency distribution over Hubble types. This
confirms the suspicion (e.g., Filippenko \& Sargent 1985; Wheeler \& Levreault
1985; Porter \& Filippenko 1987; Fransson \& Chevalier 1989) that SNe~Ibc have
massive progenitors.}

\item{The small number of data on SNe~IIn in Table 2 suggest that these objects
have a frequency distribution over Hubble type that does not differ greatly
from that of ordinary SNe~II.  In the subsequent discussion, data on SNe~II and
SNe~IIn will therefore be combined and referred to as ``SNe~II.''}

\item{Inspection of the data in Table 2 confirms that SNe~II are most common in
late-type galaxies, whereas SNe~Ia occur more frequently in galaxies of earlier
types. A Kolmogorov-Smirnov test shows that there is only a 0.8\% probability
that these two types of SNe were drawn from the same frequency distribution
over Hubble types. These data are consistent with the widely held view (e.g.,
Filippenko 1997) that the progenitors of SNe of type II are associated
with a younger and more massive population than is the case for the progenitors
of SNe~Ia. However, a possible exception is the SN~IIn 2001I, which occurred in
the E2 galaxy UGC 2836. Such a young massive object would not have been
expected among the old stellar populations in an elliptical galaxy. For
example, Filippenko (1997) writes that ``SNe~II, Ib, and Ic have never been
seen in elliptical galaxies and rarely if ever in S0 galaxies."  It would
clearly be great interest to obtain deep images to check
the E2 classification of UGC 2836. It is also possible that SN 2001I actually
occurred in a late-type dwarf galaxy near UGC 2836.}

\end{enumerate}

\section{Conclusions}

   Homogeneous Hubble-type classifications (on the DDO system) are given for
the host galaxies of the 177 SNe that have been discovered since 1997 by
LOSS. Almost all of these SNe have homogeneous spectral classifications.
The data strengthen the existing consensus that SNe~Ia occur in galaxies of all
spectral types, but that SN~Ib, Ic, and II preferentially occur in galaxies of
later type that contain a significant population of young massive stars.

\acknowledgments

One of us (SvdB) wishes to thank Chris Pritchet for comments on an early draft
of this paper and Marcin Sawicki for a search of data bases containing the
galaxy UGC 2836.  The work of A.V.F.'s group at U. C. Berkeley is supported by
the National Science Foundation grant AST-9987438, as well as by the Sylvia and
Jim Katzman Foundation. KAIT was made possible by generous donations from Sun
Microsystems, Inc., the Hewlett-Packard Company, AutoScope Corporation, Lick
Observatory, the National Science Foundation, the University of California, and
the Katzman Foundation.  A.V.F. is grateful to the Guggenheim Foundation for a
Fellowship.

\newpage

\renewcommand{\arraystretch}{0.75}

\begin{deluxetable}{llllll}
\tablecaption{Classifications of SN Host Galaxies}
\label{1}
\tablehead{
\colhead{SN} & \colhead{Galaxy} & \colhead{DDO Type} & 
\colhead{SN Type} & \colhead{V (km/s)} & \colhead{Remarks}
}
\startdata
1997bs &NGC 3627 &Sc &IIn &727&a \\
1998W &NGC 3075 &Sc II-III: &II &3582 &\\
1998Y &NGC 2415 &S &II &3784 &\\
1998bm &IC 2458 &Pec &II &1534&\\
1998bn &NGC 4462 &Sb: &Ia &1792 &b\\
1998cc &NGC 5172 &Sc I-II &Ib &4030&\\
1998cu &IC 1525 &SBb II &II &5022 &\\
1998de &NGC 252 &Sab &Ia-pec &4990 &\\
1998dh &NGC 7541 &Sc pec &Ia &2678 &b,c\\
1998dj &NGC 788 &E/Sa &Ia &4078 &\\
1998dk &UGC 139 &Sb &Ia &3963 &b \\
1998dl &NGC 1084 &Sc II: &II &1406 &b\\
1998dm &MCG -01-04-44 &Sbc &Ia &1968 &c\\
1998dt &NGC 945 &SBbc I &Ib &4480 &b\\
1998dx &UGC 11149 &E2 &Ia &14990 &\\
1998eb &NGC 1961 &Sbc pec &Ia &3934 &\\
1998ef &UGC 646 &Sb &Ia &5319 &\\
1998en &UGC 3645 &Sc &II &6385 &c\\
1998es &NGC 632 &Sab &Ia-pec &3168&\\
1998fa &UGc 3513 &Sb: &IIb &7358&\\
1998fe &NGC 6027D &? &? &19809 &d\\
1999A &NGC 5874 &Sc I &II &3128 &\\
1999ac &NGC 6063 &Sbc &Ia-pec &2848\\
1999bg &IC 758 &SBbc II: &II &1275 &\\
1999bh &NGC 3435 &S(B?)b: &Ia &8400 &\\
\enddata
\tablenum{1}
\tablenotetext{a} {Dusty.}
\tablenotetext{b} {Digital IIIaJ image not yet available.}
\tablenotetext{c} {Edge-on.}
\tablenotetext{d} {Possible merger.}
\tablenotetext{e} {Conflicting classifications.}
\tablenotetext{f} {Probable dwarf.}
\tablenotetext{g} {Too distant to classify.}
\tablenotetext{h}{It would be of interest to obtain large-scale images
   of the host of SN IIn 2001I to check if this 
   object is indeed an elliptical galaxy.}
\tablenotetext{i} {$z = 0.095$.}
\tablenotetext{j} {Possibly intergalactic.}

\end{deluxetable}

\begin{deluxetable}{llllll}
\tablecaption{(continued)}
\tablehead{
\colhead{SN} & \colhead{Galaxy} & \colhead{DDO Type} & 
\colhead{SN Type} & \colhead{V (km/s)} & \colhead{Remarks}
}
\startdata
1999br &NGC 4900 &S(B)c III: &II &969&\\
1999bu &NGC 3786 &Sbc t &Ic &2678&\\
1999bw &NGC 3198 &Sc III &IIn &663&\\
1999bx &UGC 11391 &Merger &II &4545 &d\\
1999by &NGC 2841 &Sa &Ia-pec &638&\\
1999bz &UGC 8959 &Sc I: &Ic &25362&\\
1999cd &NGC 3646 &Sc pec I &II &4248&\\
1999ce &Anonymous&E/Sa: &Ia &23400&\\
1999cl &NGC 4501 &Sb I &Ia &2281&\\
1999co &Anonymous &? &II &9260&\\
1999cp &NGC 5468 &Sc II &Ia &2845 &b\\
1999cq &UGC 11268 &S(B)bc &Ib/c &7890 &c\\
1999cw &MCG -01-02-01 &Sab: &Ia-pec &3725 &b\\
1999da &NGC 6411 &E3/Sa &Ia-pec &3690&\\
1999dg &UGC 9758 &E1/cD &Ia &6535&\\
1999dh &IC 211 &Sc II &II &3256 &\\
1999dk &UGC 1087 &Sb? &Ia &4485 &e\\
1999do &MCG +05-54-3&E2/Sa &Ia &6576&\\
1999dp &UGC 3046 &Sc: &II &4699&\\
1999dq &NGC 976 &Sab &Ia-pec &4295&\\
1999eb &NGC 664 &Sb II: &IIn &5425&\\
1999ec &NGC 2207 &Sc t I &Ib &2741 &b\\
1999ed &UGC 3555 &Sbc II &II &4835 &\\
1999ej &NGC 495 &S(B)a: &Ia &4114&\\
1999ek &UGC 3329 &Sc? &Ia &5253&\\
1999em &NGC 1637 &Sbc pec &II &717 &b\\
1999ew &NGC 3677 &Sa &II &7475 &\\
1999gb &NGC 2532 &Sc II: &IIn &5260&\\
1999gd &NGC 2623 &Merger &Ia &5535 &d\\
1999ge &NGC 309 &S(B?)c I &II &5662 &b\\
1999gf &UGC 5515 &cD or E/Sa &Ia &13293 &b\\
1999gm &PGC 24106 &Sab pec? &Ia &... &b\\
1999go &NGC 1376 &Sc II &II &4155 &b\\
1999gp &UGC 1993 &S IV &Ia-pec &8018 &c\\
1999gq &NGC 4523 &Ir IV &II &262 & f \\
1999gs &NGC 4525 &Sb II &? &1172&\\
2000A &MCG +01-59-8 &E:2 &Ia &8760&\\
2000F &IC 302 &S(B)bc II &Ic &5904&\\
2000H &IC 454 &S(B)bc: &IIb &3945&\\
2000N &MCG -02-34-5 &SBb I &II &3990 &b\\
2000Q &Anonymous&E3 &Ia-pec? &6000&\\
2000bg &NGC 6240 &Merger &IIn &7339 &b,d\\
\enddata
\tablenum{1}
\end{deluxetable}

\begin{deluxetable}{llllll}
\tablecaption{(continued)}
\tablehead{
\colhead{SN} & \colhead{Galaxy} & \colhead{DDO Type} & 
\colhead{SN Type} & \colhead{V (km/s)} & \colhead{Remarks}
}
\startdata
2000bs &UGC 10710 &S IV &II &8387&\\
2000cb &IC 1158 &Sbc III-IV &II &1927&\\
2000cc &CGCG 140-014 &Sbc &? &11073&\\
2000cg &UGC 10121 &Sbc &II &8830&\\
2000ch &NGC 3432 &S III-IV &IIn &616 &c\\
2000cn &UGC 11064 &Sc II: &Ia &7043&\\
2000cp &PGC 57064 &Sb+St &Ia &10254 &c\\
2000cq &UGC 10354 &Sbc II: &II &8946 &\\
2000cu &ESO 525-G004&S(B)b: &Ia &... &\\
2000cw &MCG +5-56-7 &Sb(p?)II: &Ia &9034&\\
2000cx &NGC 524 &E1/Sa &Ia-pec &2421&\\
2000da &UGC 5 &Sb II: &II &7271 &b\\
2000dc &ESO 527-G019 &Sb pec &II &3117 &b\\
2000dd &MCG -04-48-1&? &Ia &... & g\\
2000dg &MCG +01-1-29&Sb &Ia &1542&\\
2000dj &NGC 735 &Sb II &II &629&\\
2000dk &NGC 382 &E1 &Ia &5228&\\
2000dm &UGC 11198 &S0/Sbc &Ia &4507&\\
2000dn &IC 1468 &Sab &Ia &9613 &b\\
2000dp &NGC 1139 &S(B)b &Ia &10350 &b\\
2000dq &MCG +00-6-43&Sa &II &12687 &b\\
2000dr &IC 1610 &SB0/a &Ia &5635 &b\\
2000dt &UGC 3411 &Sc I &Ib &6798 &\\
2000du &UGC 3920 &Sc pec II: &II &8520&\\
2000dv &UGC 4671 &Sa: t &Ib &4053&\\
2000el &NGC 7290 & Sab & II &2905 & \\
2000eo &MCG -2-9-3&Sbc: &IIn&3102&\\
2000ex &MCG -5-9-22&Sbc: II &II&4093\\
2000ey&IC 1481&Sb: pec & Ia &6118&b\\
2000fa&UGC 3770&S&Ia&6378&\\
2001A &NGC 4261 &E2 &Ia &2238&\\
2001D &IC 728 &S(B?)b &II &8499 &b\\
2001E &NGC 3905 &S(B)bc I &Ia &5774 &b\\
2001F &IC 867 &S(B)bc II-III &Ia &6870&\\
2001I &UGC 2836 &E2 &IIn &4963 &h \\
2001J &UGC 4729 &SB III-IV &II &3900 &b \\
2001L &MCG -01-30-1&Sc: III-IV: &Ia &4567 &b,c\\
2001M &NGC 3240 &Sc II-III &Ic &3584 &b\\
\enddata
\end{deluxetable}
\tablenum{1}
\begin{deluxetable}{llllll}
\tablecaption{(continued)}
\tablehead{
\colhead{SN} & \colhead{Galaxy} & \colhead{DDO Type} & 
\colhead{SN Type} & \colhead{V (km/s)} & \colhead{Remarks}
}
\startdata
2001N &NGC 3327 &Sab &Ia &6303 & \\
2001P &NGC 3947 &SBb II &Ia-pec &6197&\\
2001Q &UGC 6429 &Sbc II &II &3726&\\
2001R &NGC 5172 &Sb I-II &II &4030&\\
2001Y &NGC 3362 &Sc I-II &II &8290&\\
2001Z &IC 3528 &Sc: &II &13828&\\
2001ab &NGC 6130 &Sc I: &II &5137&\\
2001ac &NGC 3504 &SBb II &IIn &1534&\\
2001ae &IC 4229 &SBb &II &6984 &b\\
2001af &MCG -04-24-1&SBbc II &II &8799 &b\\
2001ai &NGC 5278 &Sc t I: &Ic &7541&\\
2001aj &UGC 10243 &SBb II-III: &II &7937&\\
2001ay &IC 4423 &Sb III: &Ia &9067&\\
2001bp &Anonymous&? &Ia &i & g\\
2001bs &UGC 10018 &SBb II-III &Ia &8750&\\
2001cg &IC 3900 &E2/S0 &Ia-pec &7115&\\
2001ch &MCG -01-54-1&S IV &Ic &2931 &c\\
2001ci &NGC 3079 &Sbc III &Ic &1125 &c,f\\
2001cj &UGC 8399 &SBb II-III &Ia &7265&\\
2001ck &UGC 9425 &St &Ia &10408 &\\
2001cl &NGC 7260 &SBb I &II &4901 &\\
2001co &NGc 5559 &Sb III: &? &5166 &c\\
2001cp &UGC 10738 &Sb &Ia &6716 &c\\
2001cx &UGC 12266 &SBb &II &4817&\\
2001cy &UGC 11927 &Sb II: &II &4478 &\\
2001da &NGC 7780 &S(B?)b &Ia &5155&\\
2001df &MCG -04-51-5&Sc pec &II &9436 &b\\
2001dj &NGC 180 &SBb II: &II &5281&\\
2001dl &UGC 11725 &S IV: &Ia &6204 &c\\
2001dm &NGC  749 &SBab III: &Ia &4361 &b\\
2001do &UGC 11459 &Sb III-IV &II &3124&\\
2001dq &IC 1222 &S(B)bc I-II &Ic &9224&\\
2001ds &UGC  1654 &Sa &Ia &10869&\\
2001dt &UGC 12558 &Sc: &Ia &8945&\\
2001dx &PGC 63222 &Ir III-IV &II &5704 &c \\
2001ea &MCG +05-54-3&Sa &II &9566&\\
2001ec &PGC 74077 &Sa &Ia &...&\\
2001eh &UGC 1162 &SBb I &Ia &11117 &\\
2001ei &Anonymous& &Ia-pec &... & j \\
\enddata
\end{deluxetable}
\begin{deluxetable}{llllll}
\tablecaption{(continued)}
\tablehead{
\colhead{SN} & \colhead{Galaxy} & \colhead{DDO Type} & 
\colhead{SN Type} & \colhead{V (km/s)} & \colhead{Remarks}
}
\startdata
2001em &UGC 11794 &Sab III-IV: &Ic &5844&\\
2001en &NGC 523 &Merger? &Ia &4758&d\\
2001ep &NGC 1699 &Sbc: &Ia &3901 &b\\
2001eq &PGC 70417 &S &Ic &7524&\\
2001es &Anonymous&? &Ia &... & g\\
2001et &MCG -03-51-9&Sc &II &... &b\\
2001ev &UGC 2653 &Sa: &II &6939&\\
2001ez &PGC 17642 &Sa: &II &3872 &c\\
2001fa &NGC 673 &Sc I &IIn &5182&\\
2001fc &UGC 11683 &Sbc &II &5025&\\
2001fd &UGC 11957 &Sb &II &5182 &b\\
2001ff &UGC  4685 &Sa pec &II &3978 &\\
2001fh &PGC 66592 &Sb? &Ia-pec &3894&\\
2001fu &MCG -03-23-1&Sa &Ia &1730&b\\
2001fx &IC 5345&Sa: &Ib &8021 &b\\
2001fy &UGC 11922&S &II &...&\\
2001ic &NGC 7503 &E1 &Ia &13262&\\
2001id &UGC 12424 &Sc: &II &10622&\\
2001ii &UGC 444 &Sab &Ic &10664&\\
2001ir &MCG -02-22-2&Sc &IIn &5900 &c\\
2001is &NGC 1961 &Sc pec &Ib &3934&\\
2002C &IC 3376 &S(B)a II-III &II &7165&\\
2002D &NGC 594&Sa II: &II &5413&\\
2002E &NGC 4129&Sb III-IV &II &1174 & b,c\\
2002F &UGC 2885&Sbc &II &5802&\\
2002G &Anonymous&E/Sa &Ia &10114&\\
2002H &MCG -02-35-1&E1 &Ia &6679 &b\\
2002I &IC 4229 &SBb &Ia &6984 &b\\
2002J &NGC 3464 &S(B?)b &Ic &3729 &b\\
2002ao &UGC 9299 &S IV &Ic &1552 &b \\
2002aq &MCG -01-7-35&SBab &II &... &b \\
2002ar &NGC 3746 &SB0/a III: &Ia &9022 &\\
2002as &UGC 3418 &SBb II &II &6743 &\\
\enddata
\tablenum{1}
\end{deluxetable}

\begin{deluxetable}{llllll}
\tablenum{2}
\tablecaption{
Galaxy Classification and Supernova Type}
\tablehead{
\colhead{Galaxy type} & \colhead{Ia}  &\colhead{Ia-pec} &
\colhead{Ibc} & \colhead{II} & \colhead{IIn}
}
\startdata
E    &       7 &      1   &     0   &    1  &    1\\
E/Sa &       5 &      2   &     0   &    0  &    0\\
Sa   &       4 &      1   &     2   &    8  &    0\\
Sab  &       4 &      4   &     2   &    2  &    0\\
Sb   &      19 &      2   &     2   &   17  &    2\\
Sbc  &       5 &      1   &     6   &   12  &    1\\
Sc   &       6 &      0   &     7   &   18  &    4\\
Ir   &       0 &      0   &     0   &    2  &    0\\
\enddata
\end{deluxetable}
\end{document}